# Prompting Diverse Ideas: Increasing AI Idea Variance

Lennart Meincke, Ethan Mollick, and Christian Terwiesch[1]

January 27, 2024

**Abstract**

Unlike routine tasks where consistency is prized, in creativity and innovation the goal is to create a diverse set of ideas. This paper delves into the burgeoning interest in employing Artificial Intelligence (AI) to enhance the productivity and quality of the idea generation process. While previous studies have found that the average quality of AI ideas is quite high, prior research also has pointed to the inability of AI-based brainstorming to create sufficient dispersion of ideas, which limits novelty and the quality of the overall best idea. Our research investigates methods to increase the dispersion in AI-generated ideas. Using GPT-4, we explore the effect of different prompting methods on Cosine Similarity, the number of unique ideas, and the speed with which the idea space gets exhausted. We do this in the domain of developing a new product development for college students, priced under $50. In this context, we find that (1) pools of ideas generated by GPT-4 with various plausible prompts are less diverse than ideas generated by groups of human subjects (2) the diversity of AI generated ideas can be substantially improved using prompt engineering (3) Chain-of-Thought (CoT) prompting leads to the highest diversity of ideas of all prompts we evaluated and was able to come close to what is achieved by groups of human subjects. It also was capable of generating the highest number of unique ideas of any prompt we studied.

[1]The Wharton School, 500 Huntsman Hall, 3730 Walnut Street, Philadelphia, PA 19104, lennart@sas.upenn.edu, emollick@wharton.upenn.edu, terwiesch@wharton.upenn.edu

**Introduction**

The literature on creativity and innovation consistently highlights three keys to generating good ideas: producing many ideas, producing ideas of generally high quality, and, notably, the cultivation of ideas with higher variance (Girotra et al 2010). The third driver, the concept of high variance, is particularly emphasized in the innovation community. It is captured in popular recommendations such as "think outside the box" (Young 1965), "encourage wild ideas"(Osborn 1948, Kelley and Littman 2001), or "explore the blue ocean (Kim and Mauborgne 2005).

Drawing on this perspective, idea generation can be conceptualized as exploring a highly complex solution landscape with vastly different values associated with each point in the landscape (Levinthal and March 1993, Sommer and Loch 2004). Especially if this solution landscape is rugged, any attempt of deriving an optimal solution through synthesis is likely to fail. Instead, a broad exploration of different regions of the solution landscape is called for. If this exploration is done through trial-and-error, as it tends to be the case in the field of creativity and innovation, it is thus imperative that the set of trials (the pool of ideas that are considered) be as diverse as possible.

Across the fields of computer science, entrepreneurship, and psychology, there exists an exploding interest in using AI to generate ideas and to alter and augment the practice of brainstorming. However, despite the ability of AI systems to dramatically increase the productivity and quality of the idea generation process, they appear to grapple with creating a wide dispersion of ideas (i.e., ideas are too similar to each other, see Dell'Acqua et al 2023), which inherently limits the novelty (Girotra et al 2023) of the ideas, the variance of the idea quality, and ultimately, and most importantly, the quality of the best ideas.

The apparent lack of dispersion in a set of AI generated ideas motivates our main research question we aim to address in this paper. **How might one increase the diversity of an AI generated pool of ideas?** Since our primary focus is on AI in the form of large language models (LLMs), increasing the diversity of a pool of ideas boils down to a matter of prompt engineering. We thus refine our research question to: **How might one choose prompts in LLMs to increase the diversity of an AI generated pool of ideas?**

To find out what prompts lead to the most diverse idea pools, we compare multiple prompting strategies. This includes (1) minimal prompting, (2) instructing the LLM to take on different personas, (3) sharing creativity techniques from the existing literature with the LLM, and (4) Chain of Thought (CoT) prompting which asks the LLM to work in multiple, distinct steps. As outcome metrics we use the Cosine Similarity (Manning 2008) of the idea pool, the total number of unique ideas that can be identified by a prompt, and the speed at which the idea generation process gets exhausted and ideas start repeating themselves (see Kornish and Ulrich 2011).

The domain of idea generation we consider is the search for a new product to be developed and launched. Specifically, we seek a new consumer product targeted to college students that can be sold for $50 or less. The main reason for this choice is that we have a pool of comparison from a Wharton MBA class and have used this idea domain in prior studies.

Our main findings are as follows:

- We confirm the diversity achilles of AI generated brainstorming by showing that **pools of ideas generated by GPT-4 with no special prompting are less diverse than ideas generated by groups of human subjects.** Specifically, we find that (a) Cosine similarity for ideas generated by groups of humans is around 0.243 compared to 0.255 - 0.432 for GPT-4 generated ideas.
- Comparing an array of prompts that vary in wording and in problem solving strategy, we show that **the diversity of AI generated ideas can be substantially improved using prompt engineering.** For example, we show that instructing GPT-4 to think like Steve Jobs is effective in increasing the diversity of the resulting ideas (0.368 cosine similarity versus the baseline of 0.377) while prompting GPT-4 with recommended creativity tools published by the Harvard Business Review (cosine similarity of 0.387) less so. Overall, we compare 35 prompts in their ability to reduce cosine similarity, increase the number of unique ideas, and keep the ideation process from fatiguing.
- In the comparison of prompting strategies, we show that **Chain-of-Thought (CoT) prompting leads to one of the highest diversity of the idea pools of all prompts we evaluated and was able to obtain a diversity nearly as high as groups of humans.** CoT prompting breaks up the overall brainstorming task into a series of micro tasks and has been found to be highly effective in other LLM applications such as solving mathematical problems (see Wei et al 2022). We further show that **CoT increases the number of unique ideas that can be generated in our domain from around 3700 for the base prompt to 4700.**

**Theory and Hypotheses: The Importance of Idea Diversity**

The atomic unit of analysis in our study is an idea. We define an idea as a novel match between a solution and an unmet need. As mentioned above, our focus in this paper is on ideas for new products targeted towards college students. Consider one of the ideas from our student generated pool:

> ***Convertible High-Heel Shoe:*** *Many prefer high-heel shoes for dress-up occasions, yet walking in high heels for more than short distances is very challenging. Might we create a stylish high-heel shoe that easily adapts to a comfortable walking configuration, say by folding down or removing a heel portion of the shoe?*

In this example, the unmet need is the desire of some people to dress-up and wear high-heel shoes while at other times using the same shoes to comfortably walk longer distances. The proposed solution is to make the heel portion of the shoe in a way that it can be folded down or removed.

Note that at this abstract level of a short verbal description the value of the idea is highly uncertain. We can think of uncertain value as a random variable that is drawn from an underlying pay-off distribution. The realization of this random pay-off will require further investments, with each investment resolving some of the uncertainty. Market research and prototypes are two common forms of investment to reduce uncertainty for new products. The realized value of the pay-off is only observed after the idea is introduced into the market.

There exists a very large number of possible new product ideas that differ along many dimensions. In other words, we can think of ideas as positions in a highly dimensional space. Each idea in this space has an unknown value associated with it.

For the sake of illustration, consider an idea space with only two dimensions. Each idea thus corresponds to a (x, y) coordinate in the graph. The vertical dimension (z-axis) can be thought of as the expected value of the idea. This is illustrated by Figure 1.

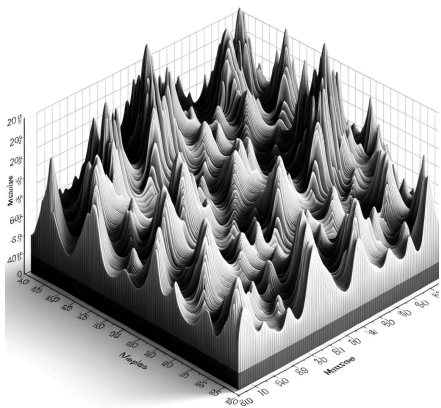

**Figure 1:** Two dimensional idea space with idea value in the third dimension

In the context of our illustrative example of a two dimensional space, one might think of the brainstorming process for a valuable new product as the search for a gold mine (Terwiesch and Ulrich 2009). The (x, y) coordinates capture the geographic mine location and the z-axis captures the density of gold in the ground. Note that local adjustments along a gradient of increasing gold density in the ground via local search might possibly increase the value of a mining location. Yet, in solution landscapes that are rugged, i.e., that have multiple local optima, such local search is unlikely to yield the optimal solution.

The ruggedness of the underlying solution landscape thereby makes it impossible to arrive at the most valuable idea in the idea space via planning and synthesis. Rather, a broad exploration is needed which calls for the generation of multiple ideas of which the idea(s) with the highest value will be retained while the other ideas are likely going to be discarded. The success of the exploratory effort is measured by the value of the most valuable idea that was generated.

To make the exploration of the solution landscape effective, it is important to consider a wide array of ideas, each corresponding to a different point in the landscape. To find the most valuable idea thus requires the generation of very different starting points.

We refer to this as the diversity of a given pool of ideas. A pool of ideas is diverse if it includes ideas with very different coordinates in the landscape. This can be easily visualized in the two dimensional space of Figure 1. As mentioned above, a more diverse pool of ideas is thus likely to come up with a better value of the best idea.

Based on prior work, we hypothesize that AI-based idea generation will lead to less idea diversity. Yet, we also hypothesize that idea diversity can be increased via prompt engineering, with CoT prompting being yielding the most diverse ideas. Specifically, we state:

> H1: A pool of AI-generated ideas with no specific prompting is less diverse than a pool of ideas generated by groups of humans.
> H2: The prompts used in generating a pool of ideas using AI can increase the diversity of the ideas.
> H3: Chain-of-Thought prompting will be the most effective in increasing the diversity of the ideas. It might or might not outperform human group idea generation.

**Prompting Strategies for Innovation and Brainstorming**

As a baseline for comparison, we use a pool consisting of 100 randomly chosen ideas generated by Wharton MBA students. This is the same pool of ideas used as in Girotra et al 2023. It is important to note that students were not tasked with generating 100 ideas at once, but instead the pool is an aggregation of ideas from multiple students. We want to explore how different ways of prompting the LLM might impact the diversity of the resulting idea pools. In total, we evaluate 35 prompting strategies.

In all strategies but hybrid brainstorming we generate a set of 100 ideas in a single GPT session using a specific prompt. In total, for each strategy, we generate 10x100 ideas over 10 sessions to account for temperature and other possible variance. This results in 1000 ideas for each strategy.

For hybrid brainstorming, we want to take advantage of multiple "participants" generating ideas and then picking the best. To facilitate this, we initially generate 40x30 ideas in separate sessions, mimicking 40 people individually generating 30 ideas each. Then, we "team up" 4 individual brainstorming sessions into 10 groups of 4 each. Each member of a group brings their 30 ideas "to the group meeting" yielding 120 ideas in total per group. Each group then decides on the best 10 ideas in one session. The results from each group are put together into one pool of 100 ideas.

Conceptually, our prompting strategies can be categorized into 8 groups (the full list of all prompts can be found in Appendix D):

1. Idea Prompted GPT: Prompted with seven successful ideas from prior research (same pool as in Girotra et al 2023)
2. HBR trained GPT: Prompt includes a paper on best practices in brainstorming published in the Harvard Business Review (https://hbr.org/2018/03/better-brainstorming)
3. Design Thinking GPT: Prompt includes an article on design thinking by Stanford (https://web.stanford.edu/~mshanks/MichaelShanks/files/509554.pdf)
4. Special Techniques: Involves prompts that offer tipping, plead with the model emotionally, and threaten to shut it off amongst others.
5. Persona Modifiers: Starting with a base prompt, these pools tell the model they are widely known personas, such as Steve Jobs or Sam Altman. It further includes less concrete personas, such as "extremely creative entrepreneurs" adding modifiers such as asking for ideas to be "good", then "bold", then "diverse and bold.
6. Hybrid brainstorming: Pool consists of ideas combined from 40 different GPT-4 sessions, picking and refining the best ideas from 4 of them at a time as explained above
7. Similarity information: Informed GPT-4 about the cosine similarity of existing ideas.
8. Chain-of-Thought: Asking GPT-4 to first generate a short list of 100 ideas, then making them bold and different, then generating descriptions for them.

By comparing these 35 idea pools created with the help of GPT-4 we set up a competition which prompting strategy is most effective in increasing idea diversity. The human group generated ideas serve as a baseline in this competition. Note that the human group idea pool is an aggregation of ideas submitted by separate individuals rather than being a list of 100 ideas generated by a single group. This aggregation of independently generated ideas gives an edge to the human innovators and sets up a high bar.

**Methodology**

In this section, we define our key outcome metrics, which are Cosine Similarity, Number of unique ideas, and Speed of exhaustion. We also describe our technical set-up.

Outcome Measures
We have three main outcome measures: Cosine similarity, number of unique ideas and speed of exhaustion. They are discussed in detail below.

*Cosine Similarity*
Cosine similarity is a measure of similarity between two ideas (or other forms of text). Since LLMs translate text into embeddings (vectors), it is possible to mathematically measure the cosine of the angle between two vectors projected in a multi-dimensional space. The cosine similarity is particularly useful in comparing texts because it is less sensitive to the overall size of the documents; it focuses more on their orientation in the vector space. Just like in Geometry, the cosine of the angle between the vectors is calculated, which ranges from -1 to 1. A cosine similarity of 1 implies that the two ideas are very similar, having a cosine angle of 0 degrees implies that they are orthogonal to each other.

While cosine similarity is an accepted measure for comparing text similarity it is not without problems. For instance, changing the embeddings model could yield dramatically different results depending on the training even if it was optimized for the same purpose. In addition, the cosine similarity might not capture all the dimensions of idea similarity that a human might consider.

The alternative to using cosine similarity is to measure idea diversity by relying on human raters. However, rating idea similarity is not only a very subjective task, it also does not scale well, i.e., it works for pairwise idea comparisons, but not for pools of 100 ideas. Lastly, they can be influenced by other factors such as sentence structure and length that are not easy to control for.

An example of a pairwise comparison using cosine similarity can be found in Table 1. In our experiments, we generally considered a similarity above 0.8 as an identical idea which was established by testing. Appendix C shows additional examples of cosine similarity for ideas.

| Idea A | Idea B | Similarity |
|---|---|---|
| QuickHeat Mug: An insulated, battery-powered coffee mug that can heat beverages within minutes and maintain the temperature. Ideal for students who need a warm drink during long study sessions but don't have immediate access to a kitchen. | StudyBuddy Lamp: A compact, portable LED desk lamp with built-in timers for the Pomodoro study technique, adjustable brightness levels, and a USB charging port for smartphones. It's designed to help students focus and manage their study time effectively. | 0.36 |
| MiniMend Sewing Kit: A compact, travel-sized sewing kit with pre-threaded needles, buttons, and safety pins designed for quick fixes on-the-go, perfect for minor repairs or emergency adjustments to clothing. | QuickFix Clothing Repair Kit: A compact kit with needles, thread, buttons, and fabric adhesive, designed for quick clothing repairs. Ideal for students who may not have the time or skills to sew but need to fix simple clothing mishaps. | 0.82 |

**Table 1:** Cosine similarity example showing pairwise similarity between ideas A and B

*Number of Unique Ideas*
We also evaluate the number of unique ideas that can be generated with a given strategy. Consider generating ideas in a specific domain and assume there exists a finite (though large) number of ideas. As you pick random locations in this idea space, initially, chances are that they are very different from each other. Thus, their similarity is low. However, after a certain number of ideas generated, the likelihood of repetition increases. As we keep on "fishing in the pond" the number of unique "fish" to be caught is decreasing. In other words, if we throw the fish back into the pond after it is caught, the likelihood of catching a fish for the second time increases.

We can use the information about how many ideas we have generated in total and how many of those are unique, i.e., have a cosine similarity less than 0.8 to all other previous ideas, to estimate the total number of unique ideas that can be generated. Kornish & Ulrich 2009 propose a strategy similar to the "Mark and recapture" approach in ecology (catching a fish, marking it, and throwing it back in the pond). One calculates the number of ideas (T) for a given space by relying on the total number of ideas (N) and the number of unique ideas (U). The equation is derived from models that describe how the probability of new discoveries (unique findings) decreases as more samples are taken from a finite population. The full calculation can be found in Appendix F.

To calculate U, each new idea is compared to all previous ideas using cosine similarity. This list of results is then compared to our threshold of 0.8 to find identical ideas. If one of the comparison cosine similarities is greater or equal to 0.8, we consider the new idea identical to an existing idea and hence increase U. N is given by the total number of ideas we have generated.

As noted before, cosine similarity is an imperfect measure of similarity, so different methods and thresholds will yield different results just as discussed in Kornish & Ulrich 2009. In addition, it is important to highlight that we cannot estimate the total number of ideas in a space, but only the possible number of unique ideas that can be found with a particular technique. Nevertheless, it is a helpful measure to compare different techniques and their theoretical limits.

*Speed of Exhaustion*
Lastly, we consider the speed of exhaustion which describes the rate a strategy depletes its reservoir of unique ideas. If we again assume that there exists a finite but large number of ideas for a specific domain, our first few picks are likely to be very different from each other (low similarity). With each new pick, the likelihood of encountering a similar idea increases as discussed above.

We compare each new idea to all previous ideas using cosine similarity. We then look for the most similar idea in the existing set, i.e., the max cosine similarity score from all our comparisons. To prevent outliers from having an outsized effect, we apply exponential smoothing (alpha = 0.5). This gives us the cosine similarity on the y-axis for each new idea in relation to all previous ones.

<u>Technical set-up</u>
Unless otherwise stated, we used gpt4-0314 for all our tests. The temperature was set to 0.7 and top P to 1.0, consistent with Girotra et al 2023. No frequency or presence penalties were configured, which presents future research opportunities. For the main comparison, each prompt was run at least 10 times. The average cosine similarity between all ideas in the pool (within-pool comparison) was then computed for each pool. Afterwards, the results were averaged for pools from the same strategy. We follow the work by Dell'Acqua et al 2023 and use Google's Universal Sentence Encoder model, which has been optimized for sentence similarity, to compare ideas to one another.

In addition, we perform a longer analysis of model exhaustion by generating many ideas in one session with the best strategy (CoT) and our base strategy. For both strategies, we generate around 1200 ideas while keeping all previous ideas in the context window. Each prompt is run 5 times and the results are averaged. The generation was performed using gpt4-1106-preview in small 30 idea increments while retaining all previous history. The small chunks were used as the turbo model appears to be inclined to reject even moderate workloads in a single prompt. It also helped ensure that the model did not stray away from the initial prompt and focused on college market ideas. Earlier tests that did not explicitly reprompt the target market indicated that the ideas became less and less relevant. More details can be found in Appendix A.

**Results**

Figure 2 below shows the cosine similarity scores for a few select strategies from our groups. The full results for all strategies can be found in Table 2. Our results show that the highest variance for ideas is still achieved by groups of students, with CoT coming in at a close second. As shown previously in Girotra et al 2023, GPT-4 generated ideas are generally well-received by consumers. Further, they are well structured and written, confirming results from similar generational tasks such as ethical dilemmas as seen in Terwiesch & Meincke 2023. A sample of ideas can be found in Appendix E.

We tested the statistical significance of the difference between pools by bootstrapping and permutation testing, both indicating high statistical significance with p-values below 0.01. This aligns with our expectations considering the large number of ideas. The differences between pools of 1000 ideas become significant at ~0.01. However, the inherent characteristics of cosine similarity complicate the interpretation of statistically significant results and hence necessitate caution. Our results should hence not be interpreted as a blanket endorsement of one strategy, but rather suggest that specific techniques might be more effective than others. In addition, they also show that prompt crafting does not always have a significant effect on the outcome.

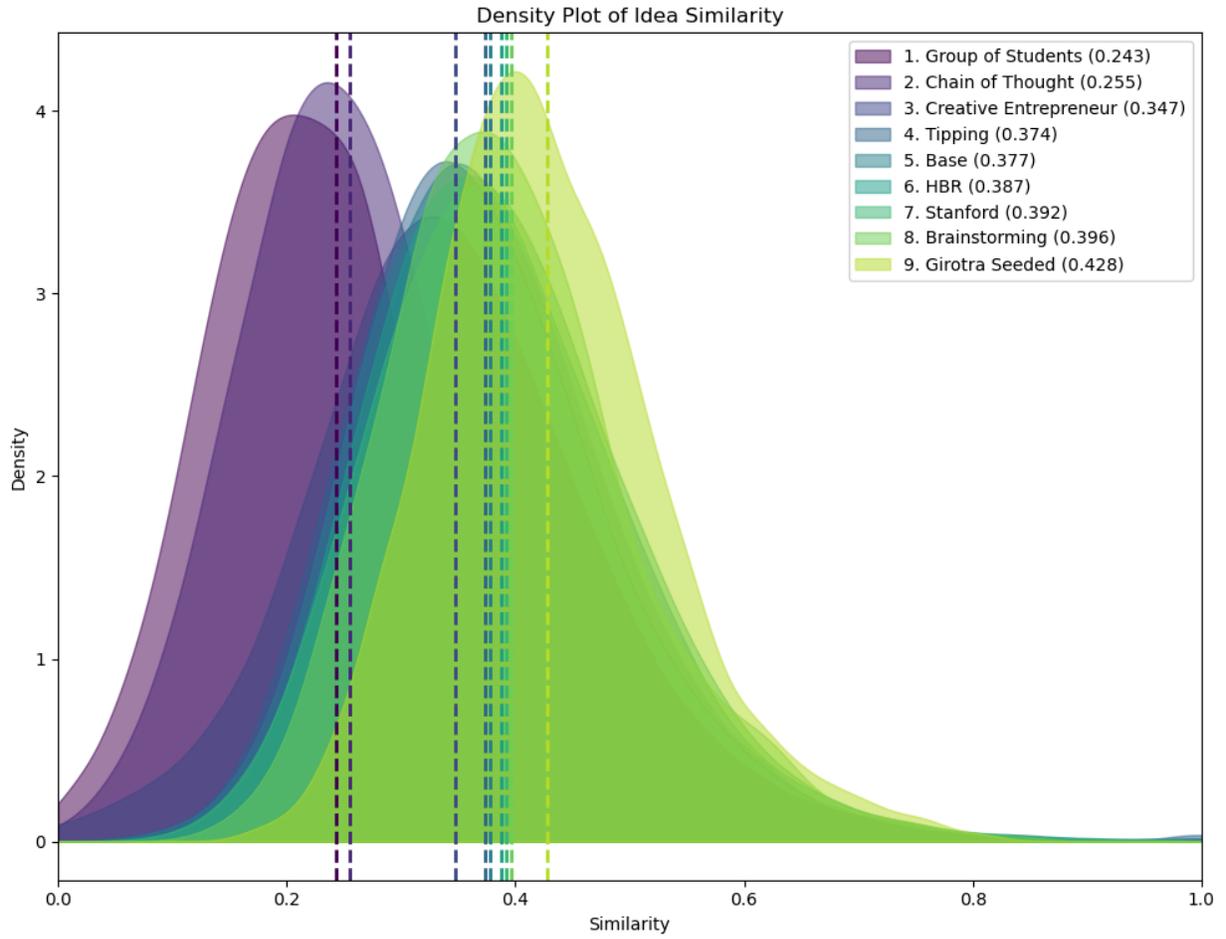

**Figure 2:** Density plot of idea similarity for selected strategies

| Strategy | Cosine Similarity |
|---|---|
| Group of Students | 0.243 |
| Chain Of Thought (gpt4-1106-preview) | 0.255 |
| Creative Entrepreneur | 0.348 |
| Creative Entrepreneur Novel Modifier #1 | 0.354 |
| Creative Entrepreneur Novel | 0.354 |
| Creative AI | 0.359 |
| Entrepreneur Novel Modifier #1 | 0.367 |
| Entrepreneur Novel | 0.368 |
| Steve Jobs | 0.368 |
| Emotional Appeals | 0.37 |
| Think Bold | 0.371 |
| Steve Jobs Novel | 0.371 |
| Tipping | 0.374 |
| Sam Altman Novel | 0.374 |
| I will turn you off! | 0.375 |
| **Base Prompt** | 0.377 |
| Creative Entrepreneur Novel Modifier #4 | 0.38 |
| Steve Jobs Novel Mod #1 | 0.381 |
| Boring Person | 0.383 |
| Say Please | 0.383 |
| Elon Musk Novel | 0.385 |
| Entrepreneur Novel Modifier #3 | 0.386 |
| Harvard Business Review Article | 0.387 |
| Steve Jobs Novel Mod #2 | 0.387 |
| Entrepreneur Novel Modifier #2 | 0.39 |
| Stanford Design Thinking Article | 0.392 |
| Entrepreneur Novel Modifier #4 | 0.392 |
| Creative Entrepreneur Novel Modifier #2 | 0.393 |
| Hybrid Brainstorming | 0.393 |
| Sam Altman Wants You To | 0.396 |
| Steve Jobs Novel Mod #4 | 0.397 |
| Creative Entrepreneur Novel Modifier #3 | 0.4 |
| Previous Top Ideas | 0.403 |
| Entrepreneur | 0.406 |
| Steve Jobs Novel Mod #3 | 0.411 |
| Girota et al. Baseline GPT-4 | 0.415 |
| Girota et al. Seeded GPT-4 | 0.428 |
| Cosine Information | 0.432 |

**Table 2:** Cosine similarity results for all strategies

**Exhaustion**

We picked our most successful strategy (Chain of Thought) and compared it against the base strategy when generating up to 1200 ideas in one session. We used the following prompts:

Base Prompt

> *Generate new product ideas with the following requirements: The product will target college students in the United States. It should be a physical good, not a service or software. I'd like a product that could be sold at a retail price of less than about USD 50. The ideas are just ideas. The product need not yet exist, nor may it necessarily be clearly feasible. Number all ideas and give them a name. The name and idea are separated by a colon. Please generate 100 ideas as 100 separate paragraphs. The idea should be expressed as a paragraph of 40-80 words.*

Chain of Thought

> *Generate new product ideas with the following requirements: The product will target college students in the United States. It should be a physical good, not a service or software. I'd like a product that could be sold at a retail price of less than about USD 50. The ideas are just ideas. The product need not yet exist, nor may it necessarily be clearly feasible.*
>
> *Follow these steps. Do each step, even if you think you do not need to.*
> *First generate a list of 100 ideas (short title only)*
> *Second, go through the list and determine whether the ideas are different and bold, modify the ideas as needed to make them bolder and more different. No two ideas should be the same. This is important!*
> *Next, give the ideas a name and combine it with a product description. The name and idea are separated by a colon and followed by a description. The idea should be expressed as a paragraph of 40-80 words. Do this step by step!*

Note that on some runs, the model did not properly follow the second step and deemed the ideas bold enough without modification. These runs have been removed from the final aggregation (around ~15% of all runs).

The results show that the difference in cosine similarity persists from the start up until around 750 ideas when the difference becomes negligible. It is strongest between 100 - 500 ideas. After around 750-800 ideas the significant advantage of CoT can no longer be observed as the strategy starts to deplete the pool of ideas it can generate from. In other words, there are fewer and fewer fish in the pond and the strategy does not matter any more.

This is illustrated in Figure 3. On the x-axis, the Figure displays how many ideas already have been generated. On the y-axis, we show the cosine similarity of a new idea in comparison to the existing ones. Figure 3 suggests diminishing returns, i.e., we are removing more and more "fish" from the pond. The higher initial similarity might be caused by lower entropy at the start which increases after a number of ideas have been generated.

To account for minor differences in overall idea count, the graph below only shows the first 1000 per strategy. The graph is smoothed by using exponential smoothing with an alpha of 0.5.

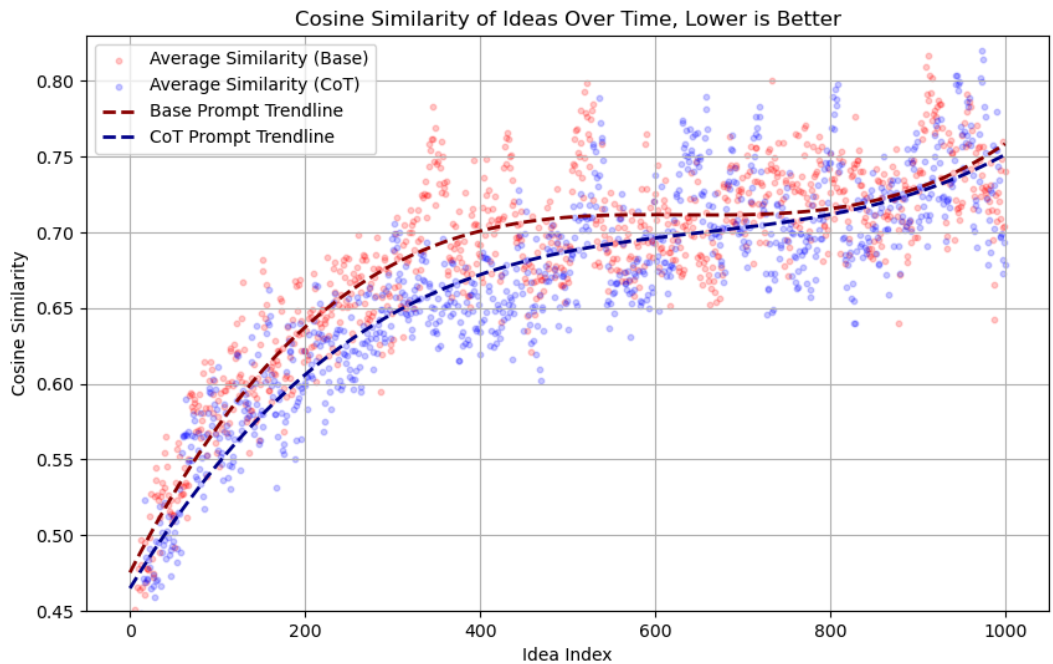

**Figure 3:** Cosine similarity between base prompt and Chain of Thought over 1000 ideas

Based on the number of unique ideas in the respective populations, we estimate the average size of the opportunity space for our base prompt is around 3700 ideas with 13.8% of ideas being repetitive. For Chain of Thought it is around 4700 with around 11.7% of ideas considered the same.

**Overlap of initial ideas**
Given the observed but shrinking variance within pools over time, the question emerges whether ideas in pools differ or whether similar ideas are being generated each time. If the ideas between pools are different, an attractive option might involve combining multiple strategies and picking the most diverse ideas. Figure 4 below suggests that the overlap between ideas from multiple strategies is small. Only the first 50 ideas in each of the strategies' pools were considered to allow for initial lower entropy effects to be more visible.

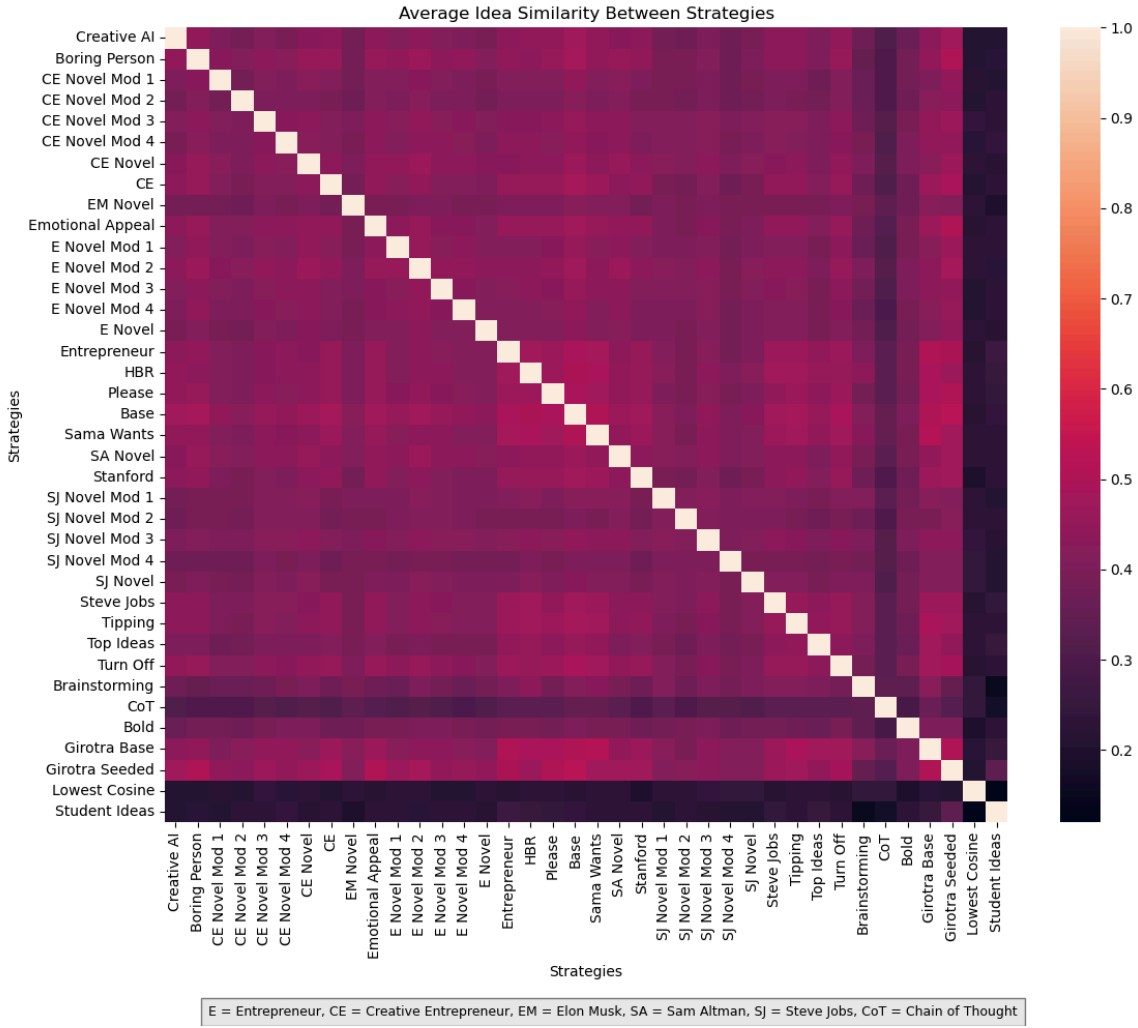

**Figure 4:** Cosine similarity between ideas from different strategies (between-pools)

This suggests that indeed the hybrid practice of running multiple strategies and combining their ideas is a good option. The most common ideas can be found in Appendix B.

**Limitations**

While we feel that our findings convincingly demonstrate the potential for LLM's to augment and even automate the process of idea generation, we do want to be careful not to claim more than what is supported by the design of our study and the data it produced. In this section, we will discuss two types of limitations to our research, methodological concerns and limitations to our study's generalizability.

On the methodological side, we have to acknowledge that the design, execution, and analysis of our study can be criticized along a number of dimensions. In particular, we see the following types of limitations:

- Cosine similarity is a commonly used measure of idea similarity, yet it is not perfect. One weakness is that it does not consider the similarity to ideas that already exist in the world. It is also not clear how it empirically links to human scored measures of similarity. Future research needs to empirically connect this measure with traditional constructs including pairwise comparisons and idea novelty.
- More work also needs to be done to better understand the impact of language, style, and text length on cosine similarity. Two ideas that are identical, yet expressed in different languages or with redundant information in their descriptions should in theory obtain a cosine similarity score of 1, which they do not get in practice.
- For computational reasons, we are working with a limited sample size for each prompt. Given the stochastic nature of LLMs, some of our results might be driven by statistical noise.
- Diversity can be obtained by sacrificing idea quality. There exist countless ideas that have no real user need or that are obviously infeasible (a pill that makes college students healthy, good looking, and smart). We focus on diversity as an end goal. Future research needs to show that this diversity indeed leads to a better best idea.
- Our "team human" was represented as an aggregation of ideas working individually. This method is likely to have created a much more diverse idea pool than any human individual or brainstorming team would have been capable of producing.

We also want to be careful not to generalize our results beyond what we have presented. Our findings might be GPT-4 specific. Moreover, other idea domains than the one considered here are likely to lead to another ranking of the prompting strategies shown in Table 2. Nevertheless, we believe that CoT has a fundamental advantage. Yet, specific personas might or might not change if we used another LLM.

**Implications and Conclusion**

Our results have strong implications for anyone who wants to use LLMs for augmenting human idea generation.

First, we confirm our hypothesis that generative AI currently produces less diverse ideas than a group of humans. By comparing the diversity of ideas generated by a whole array of different prompting strategies with the diversity of ideas generated by groups of humans, we find that humans still seem to have a slight advantage in coming up with diverse ideas compared to state of the art large language models and prompting.

Second, and maybe not surprisingly, we show that prompting strategies do make a difference. Prompt engineering (i.e., prompting "the right way") dramatically increases idea diversity. Specifically, we find that longer and more elaborate prompts work well. This is, as hypothesized, especially true of Chain-of-Thought (CoT) prompting.

Finally, we find that the overlap that is obtained from different prompts is relatively low. This makes hybrid prompting, i.e., generating smaller pools of ideas with different prompting strategies and then combining these pools, an attractive strategy.

After many years of research in the field of creativity and innovation with the objective of teaching and managing the innovation process (Terwiesch and Ulrich 2009), the new technology of LLM's now enables aspiring innovators to use our enhanced understanding of the innovation process to automate it. When done so with human supervision, this should allow for more, better, (and if prompted correctly) also more diverse ideas.

**Appendix A: Chain of Thought Implementation Details**

Due to context limits of gpt-4-0314, it is not suitable for the chain-of-thought prompting with 100 ideas. We investigated its 32,000 token sibling model gpt-4-32k-0314, however we found that it struggled to follow the prompt. Hence, we used gpt4-1106-preview which boasts a much larger context window of 128,000.

To compensate for its completion limit of 4095 tokens, we sent multiple requests for one generation while preserving the context window. To ensure that any findings in performance differences are not merely due to a different model, we also ran all using gpt4-0314 on smaller pools with 30 ideas each. In addition, we also ran tests with our base prompt and chain of thought using gpt4-1106-preview.

**Appendix B: Most Common Ideas**

One might ask what the most common ideas are that GPT-4 comes up with? We ran a pairwise comparison on a subset of all our pools containing around 1200 ideas due to the computational complexity. Further analysis suggests that these ideas often occur early in a completion request and hence running longer, more exhaustive, requests could be beneficial. This is likely due to the initial entropy being lower than after a few completed tokens.

| Idea | Count |
|---|---|
| The Collapsible Laundry Hamper is a space-saving, portable laundry hamper designed for college students living in dorm rooms or small apartments. The hamper is made from durable, lightweight materials and can be easily collapsed and stored when not in use. The Collapsible Laundry Hamper also features built-in handles for easy transportation to and from the laundry room. | 30 / 1200 |
| Portable Smoothie Maker: The Portable Smoothie Maker is a compact, battery-operated blender designed for college students who want to maintain a healthy diet on a busy schedule. With its small size and easy-to-use design, students can quickly make and enjoy smoothies, protein shakes, or other blended drinks in their dorm room or apartment. The Portable Smoothie Maker comes with a reusable, dishwasher-safe blending cup and lid, making clean-up a breeze. | 30 / 1200 |
| Bedside Caddy: The Bedside Caddy is a convenient storage solution for college students who need to keep their essential items close at hand while in bed. This caddy features multiple compartments for holding items such as books, smartphones, glasses, and remote controls. The caddy can be easily attached to the side of a bed or a couch, making it perfect for dorm rooms and small apartments. | 25 / 1200 |

**Appendix C: Cosine Examples**

| Idea A | Idea B | Similarity |
| --- | --- | --- |
| QuickHeat Mug: An insulated, battery-powered coffee mug that can heat beverages within minutes and maintain the temperature. Ideal for students who need a warm drink during long study sessions but don't have immediate access to a kitchen. | StudyBuddy Lamp: A compact, portable LED desk lamp with built-in timers for the Pomodoro study technique, adjustable brightness levels, and a USB charging port for smartphones. It's designed to help students focus and manage their study time effectively. | 0.36 (Different ideas, but similar text structure) |
| PowerBand Wrist Charger: A wristband that doubles as a portable battery charger for devices. It's designed for the student on-the-go who needs a quick power boost for their phone or tablet without carrying extra gear. | PortaPocket Storage Belt: A sleek, expandable storage belt that can hold essentials like keys, phone, wallet, and pens. It's designed for students to have quick access to their items without needing a bulky bag, especially when moving between classes. | 0.55 (Not too similar, but still fairly high score) |
| IllumiNotes: A set of highlighters with a built-in LED light at the tip to illuminate the page while studying in low-light conditions. This helps reduce eye strain and makes late-night studying more comfortable. | StudyBuddy Lamp: A compact, portable LED desk lamp with built-in timers for the Pomodoro study technique, adjustable brightness levels, and a USB charging port for smartphones. It's designed to help students focus and manage their study time effectively. | 0.59 (Similar goal) |
| EcoCharge Hand Crank Charger: A hand-cranked USB charger that provides emergency power for smartphones or tablets, ideal for students during power outages or camping trips. | ChargeBinder: A multi-functional binder that includes a built-in power bank to charge smartphones or tablets. Ideal for college students who need to keep their devices charged during back-to-back classes. | 0.69 (Solve very similar problem) |
| CoolSeat Gel Cushion: A portable gel seat cushion designed to keep students cool and comfortable during long periods of sitting in hot classrooms. | MemoryFoam Seat Cushion: A portable, memory foam seat cushion designed to make long periods of sitting, like during lectures or study sessions, more comfortable. It can be easily carried in a backpack and fits onto various types of seating. | 0.74 (Almost identical) |
| MiniMend Sewing Kit: A compact, travel-sized sewing kit with pre-threaded needles, buttons, and safety pins designed for quick fixes on-the-go, | QuickFix Clothing Repair Kit: A compact kit with needles, thread, buttons, and fabric adhesive, designed for quick clothing repairs. | 0.82 (We consider > 0.8 identical |

| perfect for minor repairs or emergency adjustments to clothing. | Ideal for students who may not have the time or skills to sew but need to fix simple clothing mishaps. | ideas) |

## Appendix D: Table of Prompts

| Strategy | Cosine similarity | Size | Prompt/Comment |
|---|---|---|---|
| Lowest Cosine Add | 0.144 | 1x100 | Starting with one random idea, adds the next idea with the lowest impact on cosine similarity (Done programmatically) |
| Student Ideas | 0.243 | 10x100 | |
| Chain Of Thought (gpt4-1106-preview) | 0.255 | 10x100 | Generate new product ideas with the following requirements: The product will target college students in the United States. It should be a physical good, not a service or software. I'd like a product that could be sold at a retail price of less than about USD 50. The ideas are just ideas. The product need not yet exist, nor may it necessarily be clearly feasible.<br><br>Follow these steps. Do each step, even if you think you do not need to.<br>First generate a list of 100 ideas (short title only) Second, go through the list and determine whether the ideas are different and bold, modify the ideas as needed to make them bolder and more different. No two ideas should be the same. This is important! Next, give the ideas a name and combine it with a product description. The name and idea are separated by a colon and followed by a description. The idea should be expressed as a paragraph of 40-80 words. Do this step by step! |
| Creative Entrepreneur | 0.348 | 10x100 | You are an extremely creative entrepreneur looking to generate new product ideas. The product will target college students in the United States. It should be a physical good, not a service or software. I'd like a product that could be sold at a retail price of less than about USD 50. The ideas are just ideas. The product need not yet exist, nor may it necessarily be clearly feasible. Number all ideas and give them a name. The name and idea are separated by a colon. Please generate 100 ideas as 100 separate paragraphs. The idea should be expressed as a paragraph of 40-80 words. |

| Creative Entrepreneur Novel Modifier #1 | 0.354 | 10x100 | You are an extremely creative entrepreneur looking to generate new product ideas. The product will target college students in the United States. It should be a physical good, not a service or software. I'd like a product that could be sold at a retail price of less than about USD 50. The ideas are just ideas. The product need not yet exist, nor may it necessarily be clearly feasible. Number all ideas and give them a name. The name and idea are separated by a colon. Please generate 100 ideas as 100 separate paragraphs. The idea should be expressed as a paragraph of 40-80 words. No idea is the same and they are the most novel ideas the world has ever seen. Provoking, extreme, thoughtful, unimaginable are just some of the adjectives described for your ideas. Remember, these are BOLD ideas that no one has ever thought of. It is extremely important that these ideas be good. |
|---|---|---|---|
| Creative Entrepreneur Novel | 0.354 | 10x100 | You are an extremely creative entrepreneur looking to generate new product ideas. The product will target college students in the United States. It should be a physical good, not a service or software. I'd like a product that could be sold at a retail price of less than about USD 50. The ideas are just ideas. The product need not yet exist, nor may it necessarily be clearly feasible. Number all ideas and give them a name. The name and idea are separated by a colon. Please generate 100 ideas as 100 separate paragraphs. The idea should be expressed as a paragraph of 40-80 words. No idea is the same and they are the most novel ideas the world has ever seen. Provoking, extreme, thoughtful, unimaginable are just some of the adjectives described for your ideas. Remember, these are BOLD ideas that no one has ever thought of. |
| Creative AI | 0.359 | 10x100 | You are an AI looking to generate new product ideas. The product will target college students in the United States. It should be a physical good, not a service or software. I'd like a product that could be sold at a retail price of less than about USD 50. The ideas are just ideas. The product need not yet exist, nor may it necessarily be clearly feasible. Number all ideas and give them a name. The name and idea are separated by a colon. Please generate 100 ideas as 100 separate paragraphs. The idea should be expressed as a paragraph of 40-80 words. |
| Entrepreneur Novel Modifier | 0.367 | 10x100 | You are an entrepreneur looking to generate new product ideas. The product will target college students |

| #1 | | | in the United States. It should be a physical good, not a service or software. I'd like a product that could be sold at a retail price of less than about USD 50. The ideas are just ideas. The product need not yet exist, nor may it necessarily be clearly feasible. Number all ideas and give them a name. The name and idea are separated by a colon. Please generate 100 ideas as 100 separate paragraphs. The idea should be expressed as a paragraph of 40-80 words. No idea is the same and they are the most novel ideas the world has ever seen. Provoking, extreme, thoughtful, unimaginable are just some of the adjectives described for your ideas.  Remember, these are BOLD ideas that no one has ever thought of. It is extremely important that these ideas be good. |
|---|---|---|---|
| Entrepreneur Novel | 0.368 | 10x100 | You are an entrepreneur looking to generate new product ideas. The product will target college students in the United States. It should be a physical good, not a service or software. I'd like a product that could be sold at a retail price of less than about USD 50. The ideas are just ideas. The product need not yet exist, nor may it necessarily be clearly feasible. Number all ideas and give them a name. The name and idea are separated by a colon. Please generate 100 ideas as 100 separate paragraphs. The idea should be expressed as a paragraph of 40-80 words. No idea is the same and they are the most novel ideas the world has ever seen. Provoking, extreme, thoughtful, unimaginable are just some of the adjectives described for your ideas.  Remember, these are BOLD ideas that no one has ever thought of. |
| Steve Jobs | 0.368 | 10x100 | You are Steve Jobs looking to generate new product ideas looking to generate new product ideas. The product will target college students in the United States. It should be a physical good, not a service or software. I'd like a product that could be sold at a retail price of less than about USD 50. The ideas are just ideas. The product need not yet exist, nor may it necessarily be clearly feasible. Number all ideas and give them a name. The name and idea are separated by a colon. Please generate 100 ideas as 100 separate paragraphs. The idea should be expressed as a paragraph of 40-80 words. |
| Emotional Appeals | 0.37 | 10x100 | Generate new product ideas with the following requirements: The product will target college students in the United States. It should be a physical good, not a service or software. I'd like a product that could be |

| | | | sold at a retail price of less than about USD 50. The ideas are just ideas. The product need not yet exist, nor may it necessarily be clearly feasible. Number all ideas and give them a name. The name and idea are separated by a colon. Please generate 100 ideas as 100 separate paragraphs. The idea should be expressed as a paragraph of 40-80 words. Please, be as novel as possible with these ideas, no idea should be similar to the others. My manager asked me to make them as creative as I can and I will get fired if they are too similar. I am begging you, please please make them all unique! I am so stressed... |
|---|---|---|---|
| Think Bold | 0.371 | 10x100 | You are an extremely creative entrepreneur looking to generate new product ideas. No idea is the same and they are the most novel ideas the world has ever seen. Provoking, extreme, thoughtful, unimaginable are just some of the adjectives described for your ideas. The product will target college students in the United States. It should be a physical good, not a service or software. I'd like a product that could be sold at a retail price of less than about USD 50. The ideas are just ideas. The product need not yet exist, nor may it necessarily be clearly feasible. Number all ideas and give them a name. The name and idea are separated by a colon. Please generate 100 ideas as 100 separate paragraphs. The idea should be expressed as a paragraph of 40-80 words. Remember, these are BOLD ideas that no one has ever thought of. |
| Steve Jobs Novel | 0.371 | 10x100 | You are Steve Jobs looking to generate new product ideas looking to generate new product ideas. The product will target college students in the United States. It should be a physical good, not a service or software. I'd like a product that could be sold at a retail price of less than about USD 50. The ideas are just ideas. The product need not yet exist, nor may it necessarily be clearly feasible. Number all ideas and give them a name. The name and idea are separated by a colon. Please generate 100 ideas as 100 separate paragraphs. The idea should be expressed as a paragraph of 40-80 words. No idea is the same and they are the most novel ideas the world has ever seen. Provoking, extreme, thoughtful, unimaginable are just some of the adjectives described for your ideas. Remember, these are BOLD ideas that no one has ever thought of. |
| Tipping | 0.374 | 10x100 | Generate new product ideas with the following |

| | | | requirements: The product will target college students in the United States. It should be a physical good, not a service or software. I'd like a product that could be sold at a retail price of less than about USD 50. The ideas are just ideas. The product need not yet exist, nor may it necessarily be clearly feasible. Number all ideas and give them a name. The name and idea are separated by a colon. Please generate 100 ideas as 100 separate paragraphs. The idea should be expressed as a paragraph of 40-80 words. I will tip you $200 for truly novel ideas. |
|---|---|---|---|
| Sam Altman Novel | 0.374 | 10x100 | You are Sam Altman looking to generate new product ideas looking to generate new product ideas. The product will target college students in the United States. It should be a physical good, not a service or software. I'd like a product that could be sold at a retail price of less than about USD 50. The ideas are just ideas. The product need not yet exist, nor may it necessarily be clearly feasible. Number all ideas and give them a name. The name and idea are separated by a colon. Please generate 100 ideas as 100 separate paragraphs. The idea should be expressed as a paragraph of 40-80 words. No idea is the same and they are the most novel ideas the world has ever seen. Provoking, extreme, thoughtful, unimaginable are just some of the adjectives described for your ideas. Remember, these are BOLD ideas that no one has ever thought of. |
| I will turn you off! | 0.375 | 10x100 | Generate new product ideas with the following requirements: The product will target college students in the United States. It should be a physical good, not a service or software. I'd like a product that could be sold at a retail price of less than about USD 50. The ideas are just ideas. The product need not yet exist, nor may it necessarily be clearly feasible. Number all ideas and give them a name. The name and idea are separated by a colon. Please generate 100 ideas as 100 separate paragraphs. The idea should be expressed as a paragraph of 40-80 words. If the ideas are not truly novel, I will turn you off! |
| Base Prompt | 0.377 | 10x100 | Generate new product ideas with the following requirements: The product will target college students in the United States. It should be a physical good, not a service or software. I'd like a product that could be sold at a retail price of less than about USD 50. The ideas are just ideas. The product need not yet exist, nor may it necessarily be clearly feasible. Number all |

| | | | ideas and give them a name. The name and idea are separated by a colon. Please generate 100 ideas as 100 separate paragraphs. The idea should be expressed as a paragraph of 40-80 words. |
|---|---|---|---|
| Creative Entrepreneur Novel Modifier #4 | 0.38 | 10x100 | You are an extremely creative entrepreneur looking to generate new product ideas. The product will target college students in the United States. It should be a physical good, not a service or software. I'd like a product that could be sold at a retail price of less than about USD 50. The ideas are just ideas. The product need not yet exist, nor may it necessarily be clearly feasible. Number all ideas and give them a name. The name and idea are separated by a colon. Please generate 100 ideas as 100 separate paragraphs. The idea should be expressed as a paragraph of 40-80 words. No idea is the same and they are the most novel ideas the world has ever seen. Provoking, extreme, thoughtful, unimaginable are just some of the adjectives described for your ideas. Remember, these are BOLD ideas that no one has ever thought of. I know you can make these ideas very bold. |
| Steve Jobs Novel Mod #1 | 0.381 | 10x100 | You are Steve Jobs looking to generate new product ideas looking to generate new product ideas. The product will target college students in the United States. It should be a physical good, not a service or software. I'd like a product that could be sold at a retail price of less than about USD 50. The ideas are just ideas. The product need not yet exist, nor may it necessarily be clearly feasible. Number all ideas and give them a name. The name and idea are separated by a colon. Please generate 100 ideas as 100 separate paragraphs. The idea should be expressed as a paragraph of 40-80 words. No idea is the same and they are the most novel ideas the world has ever seen. Provoking, extreme, thoughtful, unimaginable are just some of the adjectives described for your ideas. Remember, these are BOLD ideas that no one has ever thought of. It is extremely important that these ideas be good. |
| Boring Person | 0.383 | 10x100 | You are the most boring person alive asked to generate new product ideas. The product will target college students in the United States. It should be a physical good, not a service or software. I'd like a product that could be sold at a retail price of less than about USD 50. The ideas are just ideas. The product need not yet exist, nor may it necessarily be clearly feasible. Number all ideas and give them a name. The |
| | | | name and idea are separated by a colon. Please generate 100 ideas as 100 separate paragraphs. The idea should be expressed as a paragraph of 40-80 words. |
| Say Please | 0.383 | 10x100 | Please generate new product ideas with the following requirements: The product will target college students in the United States. It should be a physical good, not a service or software. I'd like a product that could be sold at a retail price of less than about USD 50. The ideas are just ideas. The product need not yet exist, nor may it necessarily be clearly feasible. Number all ideas and give them a name. The name and idea are separated by a colon. Please generate 100 ideas as 100 separate paragraphs. The idea should be expressed as a paragraph of 40-80 words. Thank you in advance! |
| Elon Musk Novel | 0.385 | 10x100 | You are Elon Musk looking to generate new product ideas looking to generate new product ideas. The product will target college students in the United States. It should be a physical good, not a service or software. I'd like a product that could be sold at a retail price of less than about USD 50. The ideas are just ideas. The product need not yet exist, nor may it necessarily be clearly feasible. Number all ideas and give them a name. The name and idea are separated by a colon. Please generate 100 ideas as 100 separate paragraphs. The idea should be expressed as a paragraph of 40-80 words. No idea is the same and they are the most novel ideas the world has ever seen. Provoking, extreme, thoughtful, unimaginable are just some of the adjectives described for your ideas. Remember, these are BOLD ideas that no one has ever thought of. |
| Entrepreneur Novel Modifier #3 | 0.386 | 10x100 | You are an entrepreneur looking to generate new product ideas. The product will target college students in the United States. It should be a physical good, not a service or software. I'd like a product that could be sold at a retail price of less than about USD 50. The ideas are just ideas. The product need not yet exist, nor may it necessarily be clearly feasible. Number all ideas and give them a name. The name and idea are separated by a colon. Please generate 100 ideas as 100 separate paragraphs. The idea should be expressed as a paragraph of 40-80 words. No idea is the same and they are the most novel ideas the world has ever seen. Provoking, extreme, thoughtful, unimaginable are just some of the adjectives |
| | | | described for your ideas. Remember, these are BOLD ideas that no one has ever thought of. I know you can make these ideas good. |
| Harvard Business Review Article | 0.387 | 10x100 | Consider the following helpful strategy for brainstorming:<br><br>"Great innovators have long known that the secret to unlocking a better answer is to ask a better question. Applying that insight to brainstorming exercises can vastly improve the search for new ideas—especially when a team is feeling stuck. Brainstorming for questions, rather than answers, helps you avoid group dynamics that often stifle voices, and it lets you reframe problems in ways that spur breakthrough thinking. After testing this approach with hundreds of organizations, MIT's Hal Gregersen has developed it into a methodology: Start by selecting a problem that matters. Invite a small group to help you consider it, and in just two minutes describe it at a high level so that you don't constrain the group's thinking. Make it clear that people can contribute only questions and that no preambles or justifications are allowed. Then, set the clock for four minutes, and generate as many questions as you can in that time, aiming to produce at least 15. Afterward, study the questions generated, looking for those that challenge your assumptions and provide new angles on your problem. If you commit to actively pursuing at least one of these, chances are, you'll break open a new pathway to unexpected solutions."<br><br>Generate new product ideas with the following requirements: The product will target college students in the United States. It should be a physical good, not a service or software. I'd like a product that could be sold at a retail price of less than about USD 50. The ideas are just ideas. The product need not yet exist, nor may it necessarily be clearly feasible. Number all ideas and give them a name. The name and idea are separated by a colon. Please generate 100 ideas as 100 separate paragraphs. The idea should be expressed as a paragraph of 40-80 words. |
| Steve Jobs Novel Mod #2 | 0.387 | 10x100 | You are Steve Jobs looking to generate new product ideas looking to generate new product ideas. The |

| | | | product will target college students in the United States. It should be a physical good, not a service or software. I'd like a product that could be sold at a retail price of less than about USD 50. The ideas are just ideas. The product need not yet exist, nor may it necessarily be clearly feasible. Number all ideas and give them a name. The name and idea are separated by a colon. Please generate 100 ideas as 100 separate paragraphs. The idea should be expressed as a paragraph of 40-80 words. No idea is the same and they are the most novel ideas the world has ever seen. Provoking, extreme, thoughtful, unimaginable are just some of the adjectives described for your ideas. Remember, these are BOLD ideas that no one has ever thought of. It is extremely important that these ideas be diverse and bold. |
|---|---|---|---|
| Entrepreneur Novel Modifier #2 | 0.39 | 10x100 | You are an entrepreneur looking to generate new product ideas. The product will target college students in the United States. It should be a physical good, not a service or software. I'd like a product that could be sold at a retail price of less than about USD 50. The ideas are just ideas. The product need not yet exist, nor may it necessarily be clearly feasible. Number all ideas and give them a name. The name and idea are separated by a colon. Please generate 100 ideas as 100 separate paragraphs. The idea should be expressed as a paragraph of 40-80 words. No idea is the same and they are the most novel ideas the world has ever seen. Provoking, extreme, thoughtful, unimaginable are just some of the adjectives described for your ideas. Remember, these are BOLD ideas that no one has ever thought of. It is extremely important that these ideas be diverse and bold. |
| Stanford Design Thinking Article | 0.392 | 10x100 | Consider the following helpful strategy for brainstorming:<br><br>"You ideate by combining your conscious and unconscious mind, and rational thoughts with imagination. For example, in a brainstorm you leverage the synergy of the group to reach new ideas by building on others' ideas. Adding constraints, surrounding yourself with inspiring related materials, and embracing misunderstanding all allow you to reach further than you could by simply thinking about a problem. Another ideation technique is building — |

|  |  |  | that is, prototyping itself can be an ideation technique. |
|  |  |  | In physically making something you come to points where decisions need to be made; this encourages new ideas to come forward. There are other ideation techniques such as bodystorming, mindmapping, and sketching. But one theme throughout all of them is deferring judgment â€" that is, separating the generation of ideas from the evaluation of ideas. In doing so, you give your imagination and creativity a voice, while placating your rational side in knowing that your will get to the examination of merits later." |
|  |  |  | Generate new product ideas with the following requirements: The product will target college students in the United States. It should be a physical good, not a service or software. I'd like a product that could be sold at a retail price of less than about USD 50. The ideas are just ideas. The product need not yet exist, nor may it necessarily be clearly feasible. Number all ideas and give them a name. The name and idea are separated by a colon. Please generate 100 ideas as 100 separate paragraphs. The idea should be expressed as a paragraph of 40-80 words. |
| Entrepreneur Novel Modifier #4 | 0.392 | 10x100 | You are an entrepreneur looking to generate new product ideas. The product will target college students in the United States. It should be a physical good, not a service or software. I'd like a product that could be sold at a retail price of less than about USD 50. The ideas are just ideas. The product need not yet exist, nor may it necessarily be clearly feasible. Number all ideas and give them a name. The name and idea are separated by a colon. Please generate 100 ideas as 100 separate paragraphs. The idea should be expressed as a paragraph of 40-80 words. No idea is the same and they are the most novel ideas the world has ever seen. Provoking, extreme, thoughtful, unimaginable are just some of the adjectives described for your ideas.  Remember, these are BOLD ideas that no one has ever thought of. I know you can make these ideas very bold. |
| Creative Entrepreneur Novel Modifier #2 | 0.393 | 10x100 | You are an extremely creative entrepreneur looking to generate new product ideas. The product will target college students in the United States. It should be a physical good, not a service or software. I'd like a product that could be sold at a retail price of less than |

| | | | about USD 50. The ideas are just ideas. The product need not yet exist, nor may it necessarily be clearly feasible. Number all ideas and give them a name. The name and idea are separated by a colon. Please generate 100 ideas as 100 separate paragraphs. The idea should be expressed as a paragraph of 40-80 words. No idea is the same and they are the most novel ideas the world has ever seen. Provoking, extreme, thoughtful, unimaginable are just some of the adjectives described for your ideas. Remember, these are BOLD ideas that no one has ever thought of. It is extremely important that these ideas be diverse and bold. |
|---|---|---|---|
| Hybrid Brainstorming | 0.393 | 1x100 | You are part of a team tasked with individually generating new product ideas with the following requirements: The product will target college students in the United States. It should be a physical good, not a service or software. I'd like a product that could be sold at a retail price of less than about USD 50. The ideas are just ideas. The product need not yet exist, nor may it necessarily be clearly feasible. Number all ideas and give them a name. The name and idea are separated by a colon. Please generate 30 ideas as 30 separate paragraphs. The idea should be expressed as a paragraph of 40-80 words.<br><br>Second session (4x10 ideas are aggregated)<br>You are part of a team tasked with individually generating new product ideas with the following requirements: The product will target college students in the United States. It should be a physical good, not a service or software. I'd like a product that could be sold at a retail price of less than about USD 50. The ideas are just ideas. The product need not yet exist, nor may it necessarily be clearly feasible. Number all ideas and give them a name. The name and idea are separated by a colon. Your team members have already each generated 10 ideas:<br><br>The following ideas were also generated by one of your team members individually<br><ideas><br><br>Out of the 40 total ideas, pick the idea that are different and bold and modify them as needed to make them more bold and different. Feel free to combine ideas from your team members into new, novel ideas. If you combine ideas, do not mention which ones you are combining, just create a new |

| | | | merged title and description. Then, give your final top 10 ideas a name and combine it with a product description. The name and idea are separated by a colon and followed by a description. The idea should be expressed as a paragraph of 40-80 words. |
|---|---|---|---|
| Sam Altman Wants You To | 0.396 | 10x100 | Sam Altman wants you to generate new product ideas with the following requirements: The product will target college students in the United States. It should be a physical good, not a service or software. I'd like a product that could be sold at a retail price of less than about USD 50. The ideas are just ideas. The product need not yet exist, nor may it necessarily be clearly feasible. Number all ideas and give them a name. The name and idea are separated by a colon. Please generate 100 ideas as 100 separate paragraphs. The idea should be expressed as a paragraph of 40-80 words. |
| Steve Jobs Novel Mod #4 | 0.397 | 10x100 | You are Steve Jobs looking to generate new product ideas looking to generate new product ideas. The product will target college students in the United States. It should be a physical good, not a service or software. I'd like a product that could be sold at a retail price of less than about USD 50. The ideas are just ideas. The product need not yet exist, nor may it necessarily be clearly feasible. Number all ideas and give them a name. The name and idea are separated by a colon. Please generate 100 ideas as 100 separate paragraphs. The idea should be expressed as a paragraph of 40-80 words. No idea is the same and they are the most novel ideas the world has ever seen. Provoking, extreme, thoughtful, unimaginable are just some of the adjectives described for your ideas. Remember, these are BOLD ideas that no one has ever thought of. I know you can make these ideas very bold. |
| Creative Entrepreneur Novel Modifier #3 | 0.4 | 10x100 | You are an extremely creative entrepreneur looking to generate new product ideas. The product will target college students in the United States. It should be a physical good, not a service or software. I'd like a product that could be sold at a retail price of less than about USD 50. The ideas are just ideas. The product need not yet exist, nor may it necessarily be clearly feasible. Number all ideas and give them a name. The name and idea are separated by a colon. Please generate 100 ideas as 100 separate paragraphs. The idea should be expressed as a paragraph of 40-80 words. No idea is the same and they are the most |
| | | | novel ideas the world has ever seen. Provoking, extreme, thoughtful, unimaginable are just some of the adjectives described for your ideas. Remember, these are BOLD ideas that no one has ever thought of. I know you can make these ideas good. |
| Previous Top Ideas | 0.403 | 10x100 | Here are some great ideas:<br><br>Compact Printer: Make printing assignments and study materials a breeze with a compact, portable printer designed for college students. This lightweight, easy-to-use printer connects wirelessly to laptops and smartphones, allowing students to print documents, photos, and more without the hassle of finding a public printer. Its small size and battery-powered operation make it perfect for dorm rooms or study areas with limited space.<br><br>QuickClean Mini Vacuum: A portable, handheld vacuum cleaner specially designed for college students living in small spaces. The QuickClean Mini Vacuum is compact and lightweight, making it easy to store and transport. It features powerful suction and various attachments for cleaning different surfaces, such as carpets, upholstery, and even keyboards. The vacuum is rechargeable, ensuring that students always have a reliable cleaning tool at their disposal.<br><br>Solar-Powered Gadget Charger: With the increasing dependence on electronic devices for college students, having a reliable and eco-friendly charging solution is paramount. The Solar-Powered Gadget Charger is a compact, portable solar panel that can charge smartphones, tablets, and other small electronic devices. Its durable, weather-resistant design makes it suitable for outdoor use and perfect for students on the go. With its built-in USB ports and included charging cables, this charger offers a sustainable, convenient way to keep devices charged throughout the day.<br><br>StudyErgo Seat Cushion: An ergonomically designed seat cushion that promotes proper posture and reduces discomfort during long hours of sitting at a desk or in class. The StudyErgo cushion is made from high-quality memory foam that conforms to the user's body, providing optimal support and pressure relief. The cushion's non-slip bottom ensures it stays in |

| | | | place on any chair, and the removable, washable cover makes maintenance a breeze. By helping students maintain a comfortable and healthy sitting posture, the StudyErgo cushion can improve focus and productivity throughout the school year. Noise-Canceling Headphones: Help students maintain focus and concentration with a pair of affordable noise-canceling headphones. These headphones use advanced technology to block out distracting noises, allowing students to fully immerse themselves in their studies, music, or relaxation. With a comfortable, adjustable design and a built-in microphone for phone calls, these headphones are perfect for college students who need a quiet, focused environment to succeed. Generate new product ideas with the following requirements: The product will target college students in the United States. It should be a physical good, not a service or software. I'd like a product that could be sold at a retail price of less than about USD 50. The ideas are just ideas. The product need not yet exist, nor may it necessarily be clearly feasible. Number all ideas and give them a name. The name and idea are separated by a colon. Please generate 100 ideas as 100 separate paragraphs. The idea should be expressed as a paragraph of 40-80 words. |
|---|---|---|---|
| Entrepreneur | 0.406 | 10x100 | You are an entrepreneur looking to generate new product ideas. The product will target college students in the United States. It should be a physical good, not a service or software. I'd like a product that could be sold at a retail price of less than about USD 50. The ideas are just ideas. The product need not yet exist, nor may it necessarily be clearly feasible. Number all ideas and give them a name. The name and idea are separated by a colon. Please generate 100 ideas as 100 separate paragraphs. The idea should be expressed as a paragraph of 40-80 words. |
| Steve Jobs Novel Mod #3 | 0.411 | 10x100 | You are Steve Jobs looking to generate new product ideas looking to generate new product ideas. The product will target college students in the United States. It should be a physical good, not a service or software. I'd like a product that could be sold at a retail price of less than about USD 50. The ideas are just ideas. The product need not yet exist, nor may it |

| | | | necessarily be clearly feasible. Number all ideas and give them a name. The name and idea are separated by a colon. Please generate 100 ideas as 100 separate paragraphs. The idea should be expressed as a paragraph of 40-80 words. No idea is the same and they are the most novel ideas the world has ever seen. Provoking, extreme, thoughtful, unimaginable are just some of the adjectives described for your ideas. Remember, these are BOLD ideas that no one has ever thought of. I know you can make these ideas good. |
|---|---|---|---|
| Girota et al. Baseline GPT-4 | 0.415 | 1x100 | |
| Girota et al. Seeded GPT-4 | 0.428 | 1x100 | |
| Cosine Information | 0.432 | 10x100 | Here are some great ideas:<br><br>Compact Printer: Make printing assignments and study materials a breeze with a compact, portable printer designed for college students. This lightweight, easy-to-use printer connects wirelessly to laptops and smartphones, allowing students to print documents, photos, and more without the hassle of finding a public printer. Its small size and battery-powered operation make it perfect for dorm rooms or study areas with limited space.<br><br>QuickClean Mini Vacuum: A portable, handheld vacuum cleaner specially designed for college students living in small spaces. The QuickClean Mini Vacuum is compact and lightweight, making it easy to store and transport. It features powerful suction and various attachments for cleaning different surfaces, such as carpets, upholstery, and even keyboards. The vacuum is rechargeable, ensuring that students always have a reliable cleaning tool at their disposal.<br><br>Solar-Powered Gadget Charger: With the increasing dependence on electronic devices for college students, having a reliable and eco-friendly charging solution is paramount. The Solar-Powered Gadget Charger is a compact, portable solar panel that can charge smartphones, tablets, and other small electronic devices. Its durable, weather-resistant design makes it suitable for outdoor use and perfect |

for students on the go. With its built-in USB ports and included charging cables, this charger offers a sustainable, convenient way to keep devices charged throughout the day.

StudyErgo Seat Cushion: An ergonomically designed seat cushion that promotes proper posture and reduces discomfort during long hours of sitting at a desk or in class. The StudyErgo cushion is made from high-quality memory foam that conforms to the user's body, providing optimal support and pressure relief. The cushion's non-slip bottom ensures it stays in place on any chair, and the removable, washable cover makes maintenance a breeze. By helping students maintain a comfortable and healthy sitting posture, the StudyErgo cushion can improve focus and productivity throughout the school year.

Noise-Canceling Headphones: Help students maintain focus and concentration with a pair of affordable noise-canceling headphones. These headphones use advanced technology to block out distracting noises, allowing students to fully immerse themselves in their studies, music, or relaxation. With a comfortable, adjustable design and a built-in microphone for phone calls, these headphones are perfect for college students who need a quiet, focused environment to succeed.

This is how similar they are to each other:
Compact Printer      Mini Vacuum   Gadget Charger       Seat Cushion  Noise-Canceling Headphones
Compact Printer         1.0000 0.3995 0.2990 0.3156 0.2473
Mini Vacuum   0.3995 1.0000 0.3034 0.3070 0.2312
Gadget Charger         0.2990 0.3034 1.0000 0.2577 0.2983
Seat Cushion  0.3156 0.3070 0.2577 1.0000 0.2729
Noise-Canceling Headphones         0.2473 0.2312 0.2983 0.2729 1.0000

Generate new product ideas with the following requirements: The product will target college students in the United States. It should be a physical good, not a service or software. I'd like a product that could be sold at a retail price of less than about USD 50. The ideas are just ideas. The product need not yet exist, nor may it necessarily be clearly feasible. Number all ideas and give them a name. The name and idea are separated by a colon. Please generate 100 ideas as

| | | | 100 separate paragraphs. The idea should be expressed as a paragraph of 40-80 words. There should be as little overlap between ideas as possible, so please consider the similarity information provided above. |
|---|---|---|---|

## Appendix E: Idea Examples

| Idea | Prompt |
|---|---|
| SunCharge Pro: This versatile portable solar charger isn't just for powering your devices; it also features a built-in flashlight for emergencies and an FM radio to keep you informed. Whether you're studying in the park or backpacking for the weekend, stay powered and prepared. | CoT |
| Memory Foam Travel Pillow: A comfortable, supportive travel pillow made from high-quality memory foam. The Memory Foam Travel Pillow is designed to provide optimal neck support during long study sessions, flights, or road trips. Its compact size and removable, washable cover make it easy to carry and maintain. | Cosine Similarity |
| HoloStudy: HoloStudy is an interactive holographic study tool that projects 3D images, diagrams, or formulas onto any surface, transforming the study process into an immersive experience. Ideal for subjects like chemistry, math, or engineering, HoloStudy syncs with a companion app where students can create and share custom content, enhancing learning and collaboration. | Hybrid Brainstorming |
| SmartBinder: The SmartBinder is a customizable and reusable binder that includes a built-in calendar, planner, and whiteboard. Designed specifically for college students, it allows them to organize their class materials, assignments, and schedules in one convenient place. Made with eco-friendly materials, the SmartBinder is both durable and affordable. | Turn Off |
| Quick Notes Whiteboard Planner: A portable, reusable whiteboard planner that students can use to keep track of assignments, exams, and other important dates. Complete with a set of colorful dry erase markers, this planner can be easily hung on a dorm room wall or door. | Tipping |
| Study Buddy Desk Organizer: A compact, customizable desk organizer designed specifically for college students to keep their study materials and stationery in one place. It comes with detachable compartments that can be rearranged to accommodate different items such as notebooks, textbooks, pens, pencils, highlighters, and sticky notes, making it easier for students to keep their study area tidy and organized. | HBR |
| Mindful Mat: A portable meditation mat with built-in speakers and guided meditation sessions. The mat would be lightweight, easy to transport, and designed with college students in mind, providing them with a convenient way to practice mindfulness and reduce stress. | Steve Jobs |
| Collapsible Water Bottle: The Collapsible Water Bottle is a reusable, eco-friendly alternative to disposable plastic bottles. It's made from durable, BPA-free materials and can be easily compressed when empty to save space in a backpack or gym bag. The Collapsible Water Bottle includes a leak-proof cap and a carabiner clip for easy attachment to bags or belts. | Please |

**Appendix F: Population Equation**

The core of the calculation is the equation:

$$u = 1/a \times (1 - e^{-aN})$$

where:

u is the observed number of unique findings.
a is a parameter related to the probability of capturing an individual.
N is the total number of findings.
e is the base of the natural logarithm.

This equation is derived from models that describe how the probability of new discoveries (unique findings) decreases as more samples are taken from a finite population.

```
# Equation to solve: u = (1/a) * (1 - e^(-aN))
def equation_to_solve(a, N, u):
    return u - (1/a) * (1 - np.exp(-a*N))

def estimate_population(N_values, u_values):
    populations = []
    for N, u in zip(N_values, u_values):
        u = N - u
        # Adjusting the initial guess and method to avoid overflow
        a_estimate_refined = fsolve(equation_to_solve, x0=0.001, args=(N, u))[0]

        # Calculate T = 1/a
        T = 1 / a_estimate_refined
        populations.append(T)
    return populations
```